# New AC-Powered SFQ Digital Circuits

Vasili K. Semenov, Yuri A. Polyakov, and Sergey K. Tolpygo

*Abstract*— Recent progress of Reciprocal Quantum Logic (RQL) has renewed interest in AC powering of superconductor digital circuits, which had been abandoned since the famous IBM project of 1970s. In this work we propose and demonstrate new AC-biased SFQ circuits, and search for synergy of AC and currently dominating DC biasing schemes. As the first step, we suggest an on-chip AC/DC converter capable of feeding a few DC-biased gates surrounded by their AC-biased counterparts. As the second step, we introduce and present the first successful demonstration of a new AC-powered circuit – an 8192-bit shift register with over 32,800 Josephson junctions (JJs) and JJ density of about $6 \cdot 10^5$ JJ per $cm^2$. We suggest a few niche applications for this type of AC-biased circuits, not requiring high clock rates. *E.g.*, these, scalable to millions of JJs per chip, circuits can serve as a convenient benchmark for new SFQ fabrication technology nodes, allowing the operating margins of individual cells to be extracted and, thus, "visualize" individual fabrication defects and flux trapping events. The circuit can also be developed into a mega-pixel imaging array for a magnetic field microscope.

*Index Terms*—superconductor digital electronics, circuit self-diagnostics, magnetic field microscope.

## I. INTRODUCTION

ALMOST all modern desktop computers and workstations are plugged into AC power lines, whereas the integrated circuits used in these computers are DC-biased devices powered via AC/DC converters. In contrast to this simple picture of CMOS technology, the entire history of superconductor digital circuits could be viewed as a race between AC- and DC-biasing schemes. The best known period of AC-biasing domination is associated with the famous IBM project on Josephson computer [1],[2]. After that, for about 20 years, the superconductor digital electronics was dominated by DC-biased RSFQ circuits [3]. Very recently, AC-biased RQL technology took a lead in some measures [4]-[7]. Each of the biasing schemes has its own advantages and disadvantages. In particular, DC biasing provides higher clock frequency and more flexible timing solutions. However, it suffers from the necessity to supply very large currents, growing linearly with the circuit complexity. In contrast, AC biasing dramatically reduces currents applied to superconductor integrated circuits. This advantage, however, comes at quite a high price. In particular, the distribution of multiphase, high-frequency (10 GHz to 30 GHz) bias currents between gates could be a challenging technical task. Our ultimate goal is to breed and synergize DC and AC biasing techniques. Initially, we anticipated this to be a difficult task because of highly partisan debates between proponents of DC- and AC-biased SFQ circuits. Thus, we expected to find a "high wall" separating these techniques. Fortunately, we have not found any. Instead, we discovered that data terminals of DC- and AC-biased cells could be directly connected with each other. It was a very satisfying discovery that DC and AC biasing schemes can not only coexist in the same integrated circuits, but generate a great synergy.

## II. AC/DC CONVERTER OR JOSEPHSON JUNCTION RECTIFIER

### A. DC SQUIDs as superconductor diodes

Efficient on-chip AC/DC and DC/AC converters would be the most straightforward remedy dramatically reducing the incompatibility of AC and DC biasing schemes. In fact, any Josephson junction is able to convert a DC voltage to an AC current and therefore serve as a DC/AC converter. Such converters could be quite efficient. In particular, they are inherent components of nSQUID based logic circuits [8], [9]. A possible AC/DC converter or AC rectifier is, in fact, nothing more than a well-known two-junction SQUID [10] with a constant magnetic bias $\varphi_e$.

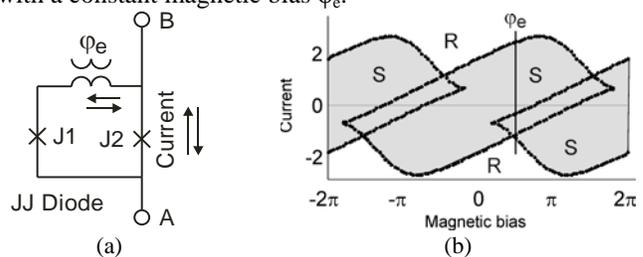

Fig. 1. Schematics (a) and state diagram (b) of a diode based on a magnetically-biased two-junction SQUID. The diagram is shown for $I_{c1}$ = 1.68 (0.21 mA), $I_{c2}$ = 1 (0.125 mA) and loop inductance $l$ = 1.55 4.09 pH).

The desired rectifying effect is illustrated in Fig. 1(b) by a conventional state diagram (see, for example [10]) that shows boundaries of static S-states with different numbers of stored flux quanta. The uppermost and lowermost pieces of the boundaries simultaneously serve as the boundaries between the static (superconducting) and dynamic (resistive) R-states. In other words, they show the variations of the positive and negative critical currents with the applied magnetic flux or the phase shift if the more convenient dimensionless terminology is used. Figure 1(b) shows that the desirable rectification corresponding, for example, to the maximum positive and the minimum negative critical currents is reached at $\varphi_e \sim \pi/2$. The required magnetic bias (phase shift $\varphi_e$) in the secondary coil of the transformer in Fig. 1(a) is created by a DC current applied to the primary coil. It can be seen now that the device is,

V.K. Semenov (corresponding author) and Yu.A. Polyakov are with the Department of Physics and Astronomy, Stony Brook University, Stony Brook, NY 11794-3800, USA; e-mail: Vasili.Semenov@StonyBrook.edu, Yuri.Polyakov@StonyBrook.edu

S.K. Tolpygo is with MIT Lincoln Laboratory, Lexington, MA 02420, USA, e-mail: sergey.tolpygo@ll.mit.edu.

indeed, a JJ-based diode - it is superconducting for one direction of an external current applied between terminals A, B in Fig. 1(a) and resistive for the opposite direction. The state boundary diagram has been calculated at parameters shown in Fig. 1 caption. The parameter values hereafter are given both in the SI and dimensionless units using 0.125 mA is a unit of current and 2.64 pH as a unit of inductance.

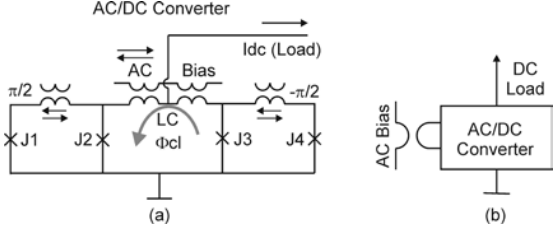

Fig. 2. Josephson junction based AC/DC converter (a) composed of two JJ-based diodes shown in Fig. 1 and its notation (b).

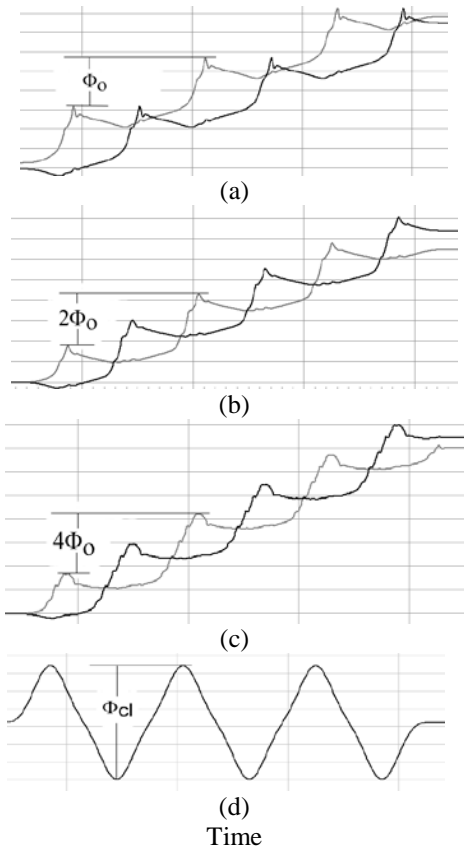

Fig. 3. Time evolution of flux flowing via Josephson junctions J2 (grey traces) and J3 (black) for a primitive (a), basic (b), and advanced (c) operation modes corresponding to $\Phi_0$, $2\Phi_0$, and $4\Phi_0$ flux jumps per each AC clock period. The mode of operation is regulated by the amplitude of clock signal $\Phi_{cl}/2$ (d). The simulations have been carried out at LC=3.74 (9.9 pH), load current Icd=0.84 (0.1 mA), and $\Phi_{cl}/2\Phi_0$ equal to 1.46 (a), 2.5 (b) and 4.0 (c). The clock signal has $\Phi_0/2$ offset in primitive mode and no offsets in the other modes. Clock frequency is about 9 GHz.

*B. AC/DC Converters*

An AC/DC converter or rectifier (Fig. 2) can be composed of two SQUIDs (JJ-based diodes) shown in Fig. 1(a). Its new component is a transformer that converts AC bias current into an AC current that flows through both SQUIDs. The direction of this current is the same for both SQUIDs, while their magnetic biases have different signs. As a result, at any direction of the AC current, one SQUID is superconducting and the other is resistive.

The proposed operation of our SFQ rectifier is supported by numerical simulations shown in Fig. 3. If the amplitude of AC clock, $\Phi_{cl}/2$, is relatively small, only one flux quantum ($\Phi_0$) passes through each junction per each clock period, as shown in Fig. 3(a). That is, the phase drop across each junction makes one full $2\pi$ jump per the AC bias (clock) period. Note that J2 phase makes $2\pi$ jumps at positive phases of AC clock, Fig. 3(d), whereas J3 makes similar jumps at negative AC phases. The number of $2\pi$ jumps per period can be increased simply by increasing the amplitude of the AC clock. For instance, Fig. 3(b) and (c) correspond to passing of, respectively, 2 and 4 flux quanta through each junction per AC period. The simulations show only 3 periods of the AC bias.

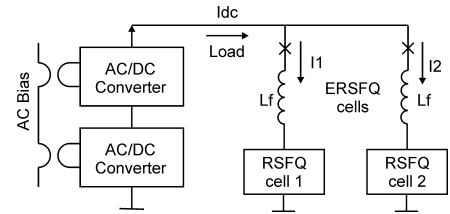

Fig. 4. AC/DC converters could be connected in series and they could feed several DC biased cells. The shown combination of RSFQ cells biased via inductances Lf and shunted Josephson-junctions is known as ERSFQ technique [11].

The output voltage of the AC/DC converter at terminal Load in Fig. 2(a) crudely "averages" voltage drops across all 4 Josephson junctions. The DC component of the voltage is defined by the AC clock rate using the Josephson frequency-to-voltage relationship with the optional integer factor (2 or 4) discussed in the previous paragraph. Undesirable AC components of the output voltage could be depressed with a low-pass filter. However, it is not always necessary to include the filter as an inherent part of the converter. For example, Fig. 4 shows the converter feeding two RSFQ cells. Large inductances *Lf* connected in series with Josephson junctions transform conventional RSFQ cells into their ERSFQ counterparts [11]. In this case, the required low-pass filters are already presented as inherent components of ERSFQ cells.

At first glance the suggested converter appears too complex to be practical. This is not so, however, because the suggested solution can eliminate a highly unpleasant mismatch between the relatively low ~12 GHz frequency provided by AC clock distribution lines [12] and much higher (100 GHz and above) routinely reported clock rates of DC-biased circuits.

The minimum DC bias voltage required for the ERSFQ cells [11] is exactly equal (through the Josephson voltage-frequency relation) to their maximum operation rate. In practical terms, this means that the primitive operation mode (Fig. 3a) of the converter provides equal ERSFQ and AC clock rates. However, we showed that the converter can also operate in a more robust basic mode (Fig. 3b) sufficient to double the ERSFQ clock rate at the same AC clock rate. The advanced operation mode (Fig. 3c) increases the ratio of ERSFQ and AC clock rates up to four.

There are two options for further increasing the clock ratio. It can be achieved by a corresponding re-optimization of the AC/DC converter. The other straightforward solution is to connect two (or more) converters in series. For example, block diagram shown in Fig. 4 provides a clock ratio of eight. That is, the figure illustrates how to run ERSFQ gates at 80 GHz rate with only 10 GHz AC clock frequency. In this case one DC-biased cell is able to make eigth logic operations during one AC clock period.

Factor 8x is a good number for hardware saving. For example, transferring data over microstrip lines with around 100 GHz rates is routinely reached with DC-biased drivers and receivers [13]-[15]. Now, the AC/DC conversion technique suggested here can be used to feed the drivers and receivers as well as simple DC-biased multiplexers and demultiplexers. As a result, it will be possible to compress data from 8 AC-biased channels into one microstrip line. Such an improvement could be vitally important for AC-biasing-based technologies as a powerful remedy against the "tyranny of wires" or, more specifically, against "tyranny of microstrip lines".

The hybrid (AC/DC) technique introduced here could be equally beneficial for once "pure" DC-biasing-based technologies suffering from too high bias currents. Indeed, in Introduction we mentioned that computers are usually AC-powered. Inside the computer, the applied relatively low AC current is converted into several larger DC currents by AC/DC converters. The AC/DC conversion proposed here could play a similar role with one essential difference. In CMOS technology one AC/DC converter is a separate "hardware box" feeding many integrated circuits. In our case, several (many) converters should be located inside each integrated circuit.

Until now the only alternative solution for reducing DC bias current has been the so-called current recycling. Theoretically, this very powerful solution originated from a very old [16] and more recent [17] works. Despite a successful demonstration [18] and numerous speculations, the current recycling technique still has not taken off. In other words, we are not aware of any off-the-shelf current recycling solutions.

## III. AC-Biased Flux shuttle

### A. Cell design

Our interest in AC biasing technique was initiated by a recent progress of RQL [4]-[7]. In particular, it was interesting to compare RQL with much older AC-biased circuits. In 1977, we suggested a so-called flux shuttle [19], [20] shown in Fig. 5(a) in its modern realization. The circuit has been re-optimized using modern CAD tools. Schematics optimization has been carried out with PSCAN/Cowboy set of tools. Layout has been matched with schematics using InductEx-FastHenry package [22],[23] and two M. Khapaev packages [24], [25].

The flux shuttle cell (Fig. 5) uses only four Josephson junctions per bit. The old cell [19], [20] was clocked by an AC current flowing directly through the cell. The new solution includes a transformer that galvanically isolates but magnetically couples the cell and the AC clock line.

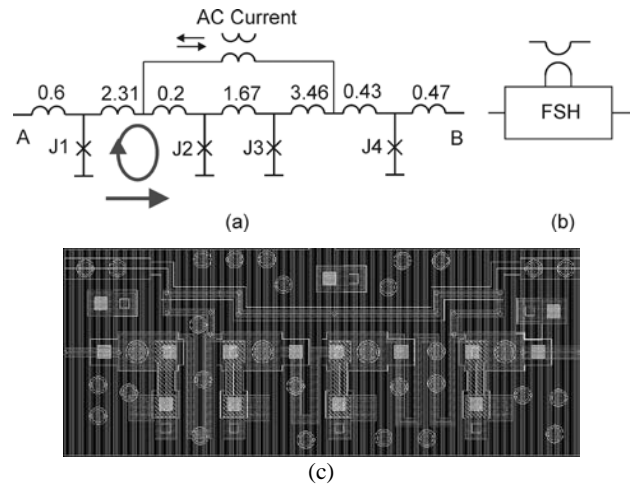

Fig. 5. Equivalent circuit (a), notation (b) and layout (c) of a one-bit cell of an AC-clocked flux shuttle. Values of inductances are shown in PSCAN units (1 PSU = 2.64 pH), critical currents of all 4 junctions are about 0.125 mA; Shown layout with 40 μm x 17 μm cell dimensions has been developed for the MIT Lincoln Laboratory fabrication process [26],[27] node SFQ3ee.

The operation of the cell is quite simple. Assume that initially a fluxon is stored between junctions J1 and J2. It jumps to the next stable position between junction J3 and J4 when AC current in the secondary coil of the transformer, applied to junctions J1 and J2, exceeds some positive threshold value, causing J2 to switch. The fluxon cannot move further because junctions J3 and J4 are negatively biased by the same AC current. During the next half of the clock period the bias current reverses its sign. Now J3 and J4 become positively biased. J4 switches when the total induced current exceeds the critical values, and Lorentz force pushes the fluxon further to the right, into the next cell.

### B. Chip design

Our most general design goal was to evaluate the quality of a new fabrication process offered by MIT Lincoln Laboratory [26], [27]. One of the metrics for such evaluation could be the maximum complexity or the maximum length of a shift register that could be successfully fabricated. To achieve this goal we designed a chip with shift registers of several different lengths (see Fig. 6), assuming that - if the longest shift register does not work - we will be able to find a complexity threshold for the currently offered fabrication technology. To extract the complexity threshold with some accuracy, we added to the registers several taps with RSFQ monitors that "visualize" the propagation of fluxons along internal cells. After the fabrication, fully operational shift registers of all implemented lengths have been found. (One can say that our search for the complexity threshold failed.)

As we mentioned earlier, we were interested in "breeding" our old DC-biased (RSFQ) cells library with the new AC-biased cells. Initially we thought that it would be a challenging task. However, the very first numerical simulations showed that data terminals of AC and DC biased cells could be connected directly with only minor or no parameter revisions. As a result, all I/O channels have been composed of our heritage collection of RSFQ cells. Of course physical layouts of all components (inductances, Josephson junctions, resistors, *etc.*) have been squeezed to satisfy the new design rules.

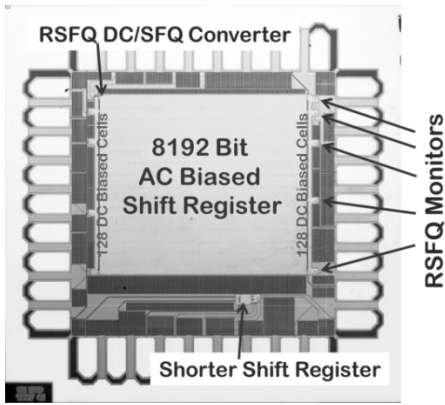

Fig. 6. Microphotograph of a 5 mm x 5 mm chip with a large (64 x 128)-bit and small (3 x 4)-bit AC-biased shift registers. Locations of DC-biased RSFQ components (input DC/SFQ and output SFQ/DC converter, *etc.*, see also Fig. 7) are shown.

The longest of the described shift registers contains 8192 flux shuttle cells (matrix of 128 cells x 64 rows) and seven intermediate I/O "taps." The circuit occupies almost all the 3 mm x 3 mm "payload" space available on a 5 mm x 5 mm chip. As mentioned earlier the 40 µm x 17 µm dimensions of the cell correspond to an impressive junction density about 600,000 JJ/cm$^2$.

Details of the connection of AC and DC biased cells are explained in Fig. 7. Rows of flux shuttle (FSH) cells are assembled from directly connected cells shown in Fig. 5. Odd rows are horizontally flipped with respect to even rows. AC clock/power lines of all FSH cells are connected in series. Simple single-junction pieces of a Josephson Transmission Line (JTL), see Fig. 7(b) and (c), connect neighboring odd and even rows of FSH cells. A tap is created by replacing the JTL with an RSFQ splitter (Splt) with fan-out two. The free output of the splitter feeds an RSFQ SFQ/DC converter. We already mentioned that all RSFQ cells are biased in parallel. Generally this is a great drawback to almost any DC-biased circuit. However, the total number of DC-biased cells in this circuit is much lower than the number of its AC-biased cells. As a result, the parallel DC biasing does not affect our particular circuit.

## IV. MEASUREMENTS

As we mentioned earlier, the distribution of very high (GHz-range) frequency signals along the chip is one of the major challenges of AC biasing. Fortunately, benchmark and yield statistics can be measured at low (*e.g.*, kHz-range) clock rates, which was utilized in this work.

The process yield is a percentage ratio of operational circuits. It is too early to report this figure because only three samples have been fully measured. The next straightforward step is the extraction of operation margins of the circuits. Comparison of the expected and extracted margins can be used to conclude the absence or presence of fabrication defects or trapped flux. Generally, four figures of merit could be extracted from AC clock margins: Positive Upper (PU), Positive Lower (PL), Negative Lower (NL), and Negative

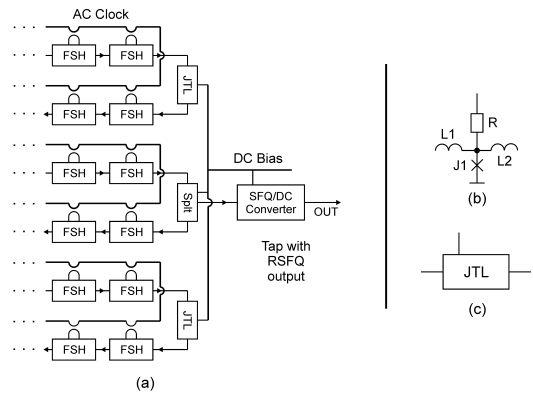

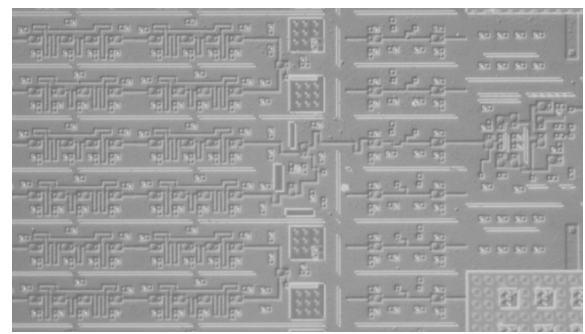

(d)

Fig. 7. A block diagram (a) and a microphotograph (d) of one fragment of the flux shuttle chip. Clock (bias) lines of flux shuttle (FSH) cells are connected in series, whereas DC power lines for RSFQ cells (JTL, Splitter (Splt) and SFQ/DC converter) are connected in parallel. Single-junction JTL sections (b), (c) are used to connect rows of FSH cells. Identical Josephson junctions with about 0.125 mA critical currents are used in FSH and JTL cells.

Upper (NU) amplitudes. Within these global margins all cells of the entire circuit are operational. However, the mentioned figures are not very informative. They tell us about the margins of only one or a few worst cells, but do not characterize the majority of cells operating with wider margins.

We have figured out how to extract margins of individual cells. Our technique is based on the assumption that cells have different margins for working with logic "1"s and "0"s. Numerical simulations have shown that at least some of the 4 types of margins are much wider for logic "0" than for logic "1". As a result, data patterns with single "1" provide a "half selection" of one particular cell, because only the cell storing "1" is sensitive to manipulations by the clock signal.

To extract one of four properties of one particular cell $k$, we should write logic "1" only into that cell; all other cells should store "0"s. We can do this at the optimal (within the global margins) positive and negative amplitudes, when all cells are operational. Then, we apply one clock pulse with altered amplitude of either its positive or negative half-period as shown in Fig. 8. If the cell works at the altered amplitude, it will shift "1" into the next position, $k + 1$. This can be checked out when we read out the whole register, using again the best clock pattern amplitude. That is, depending on if cell $k$ worked or not at the altered clock amplitude, we will find logic "1" ether in the expected position $k + 1$ or in a different position. At too low amplitude, the "1" will remain in the same position $k$. At excessive amplitude, the logic "1" will be shifted by more than one position. Both facts can be detected by finding "1" in the corresponding position in the output pattern.

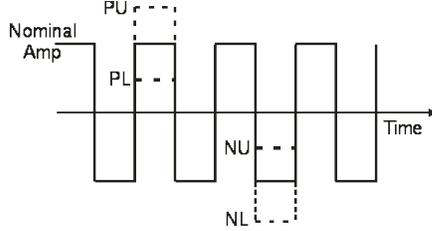

Fig. 8. Rectangular AC clock signal used in the measurements. For normal operation, all clock pulses have the same amplitude. At low clock frequencies, it is easy to intentionally change amplitude of one half-period (positive or negative) of only one clock pulse. This technique was used for measuring margins of individual cells.

To extract all four margins of cell $k$ we should repeat the experiments using all different alterations of the clock pulse number $k$. To extract properties of all cells, we should repeat the described procedure for all values of $k$. The results of realization of this algorithm are given below. We were able to dramatically accelerate the described algorithm to provide a full scan of eight thousand cells during less than one minute.

*A. Statistics of cell properties*

Figure 9(a) shows Negative Upper (NU) margin of cells in the shift register. This is a 3-dimensional plot with $z$-axis and color coding showing NU values for individual cells, and $x$, $y$-axes showing cells locations in the 128 x 64 matrix of the shift register.

The most striking feature is a sharp single-cell positive peak of NU at the 64-cell edge of the register. The next feature is a visible elevation of margins of all cells along both 128-cell edges of the register. Interior cells demonstrate quite low variations of the NU margin, within ± 0.1 mA.

These observations are illustrated by Fig. 9(b) that shows separate histograms for edges with 128 cells, edges with 64 cells, and for interior cells. A green circle positioned at about −0.5 mA corresponds to the mentioned earlier sharp peak in NU margin. The margins of cells at the edges with 128 cells, shown by blue circles, are also slightly shifted toward less negative values. For a comparison, histograms of Positive Lower (PL) margins are also shown in Fig. 9(b) at positive $x$-values. The PL histograms have a slightly different pattern showing one cell with a depressed PL margin and one cell with an increased PL margin in comparison with the dense flock of the other cells. The margins are highly reproducible with repetition of the complete measurement sequence.

The obvious question is about the nature of deviations of individual margins. They could be originated from either the fabrication process or flux trapping, or both. To resolve this, we thermally cycled the circuit above the critical temperature and repeated the measurements. The histograms have changed after the thermal cycling, certainly indicating that the margin peaks are caused by flux trapping. The histogram of the differences of NU margins before and after the thermal cycling is shown in Fig. 10.

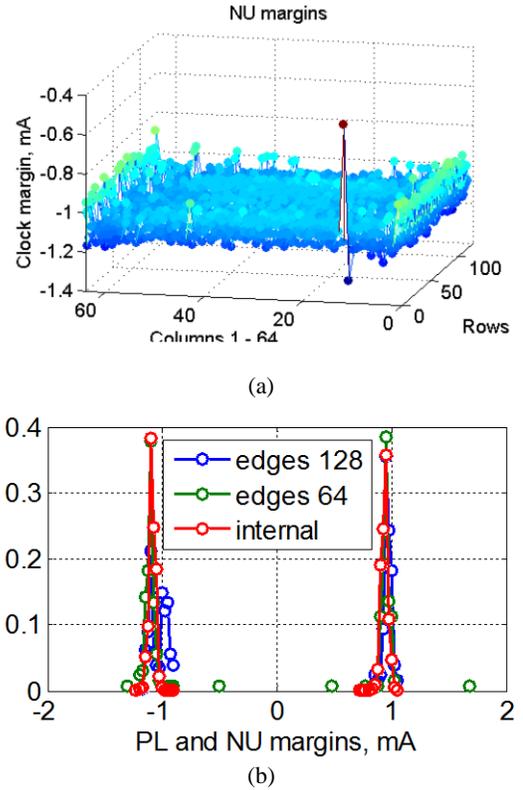

Fig. 9. Statistical properties of margins of individual cells of a (64 x 128)-bit shift register can be illustrated by a 3D plot (a) or by a histogram (b). Each color-coded point in (a) shows NU margin of the corresponding cell in the shift register.

A high peak at zero current, corresponding to no change in the margin, represents the cells that have not been affected by flux trapping. A narrow width of this peak, about ± 0.1 mA, indicates that a large number of cells have been affected insignificantly, *i.e.*, the changes of their margins are small despite the trapped flux. This type of flux trapping could be referred to as "good" in the sense that the flux was trapped where it was supposed to be trapped, - in the dedicated moats - doing no real damage to the circuit operation. In addition, we also see a "bad" flux trapping, *i.e.*, the appearance of new cells at the 64-cell edge with significantly altered margins. In Fig. 10 these cells are represented by green points remote from the main peak. Both the location of the affected cells and the amount of change in the margin became different after the thermal cycling.

We demonstrated a similar visualization technique in [28]. However, it was implemented for a DC-biased circuit, where we could not manipulate with intensity of individual SFQ clock pulses. To enable flux trapping visualization, this drawback was compensated by an intentional alteration of the circuit. As we showed above, the AC bias is much more convenient because it permits the margins of "payload" or unaltered circuits to be analyzed.

The presented flux visualization technique is a kind of self-diagnostics. It can be compared with known magnetic field microscopes. Our, approximately 20 µm, spatial resolution is not as impressive as about 1 µm resolution demonstrated in [29]. However, the scanning procedure is fast, about a minute for a 3 mm by 3 mm field, and can be easily accelerated. The resolution of so-called SHPM technique [30]-[32] used by G. Stan e*t al*. is based on a slow mechanical scanning of the magnetic probe. Besides, commercially available SHPM microscopes have small (~10 µm to 20 µm) observation fields.

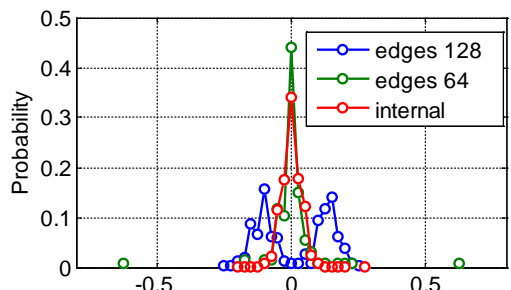

Fig. 10. Histogram of the difference of NU margins extracted before and after a thermo-cycle above $T_c$.

The margin scanning technique presented above perfectly satisfies our diagnostic needs, but it cannot be readily applied to investigate flux trapping in any superconductor integrated circuits or flux distribution in any object. This is because the shift register is intentionally shielded well from an external magnetic field and, therefore, external field coupling is a parasitic effect. It is easy to partly depress the shielding of storage loops between J1,J2 and J3,J4 pairs in Fig. 5(a) ("sensing SQUIDs") and therefore increase their expose to a magnetic field to be investigated. After such an update and with a smaller pixel (cell) size, the circuit can serve as a mega-pixel magnetic sensing array. To visualize magnetic field distribution in a different circuit or a sample, this array should be simply placed face to face with the circuit or sample under investigation, and the margin scanning should be done to generate images similar to Fig. 9(a).

## V. Conclusion

Most of presented here results were unexpected for us. We discovered that DC and AC biasing schemes can not only coexist in the same integrated circuits, but generate a great synergy.

We tried to figure out how to deal with challenging multi-GHz external (off-chip) electronics required for AC biasing schemes. Instead, we discovered a broad field of low-frequency AC-biasing applications, which does not require any special equipment.

We tried to develop a few new benchmark circuits that could be placed in a row with many other technological tests. Instead, we discovered that the same circuit can serve as a unique self-diagnostic tool for "visualizing" individual technological defects and flux trapping events. Moreover, the benchmark circuit could be transformed into a mega-pixel magnetic sensing array.

Of course, such a sequence of unexpected positive developments came at a price. In particular, some prospective logic families, for example, [33] have not been reviewed for compatibility with our synergetic AC/DC unification.

ACKNOWLEDGMENT

We would like to thank Marc Manheimer, Anna and Quentin Herr for stimulating discussions. We would like to thank V. Bolkhovsky and T. Weir for their help and W.D. Oliver, L.M. Johnson, and M.A. Gouker for their interest and support of this work.

Very special thanks go to C. Fourie, M. Volkmann, and M. Khapaev who provided intensive support for their unique software packages, and to D.V. Averin and K.K Likharev for valuable remarks.